\documentclass[manuscript]{aastex}

\slugcomment{Not to appear in Nonlearned J., 45.}

\shorttitle{Collapsed Cores in Globular Clusters}
\shortauthors{Djorgovski et al.}

\begin{document}


\title{Towards Understanding the Nature of \\
Young Detached Binary System HD 350731}


\author{F. Soydugan\altaffilmark{1} and F. Ali\c{c}avu\c{s}\altaffilmark{1}}
\affil{Department of Physics, Faculty of Arts and Sciences, \c{C}anakkale Onsekiz Mart University, TR-17100 \c{C}anakkale, Turkey}
\email{fsoydugan@comu.edu.tr}
\author{S. Bilir}
\affil{Department of Astronomy and Space Science, Faculty of Science, \.{I}stanbul University, TR-34119, University-Istanbul, Turkey}

\and

\author{E. Soydugan\altaffilmark{1}, \c{C}. P\"{u}sk\"{u}ll\"{u}\altaffilmark{1} and T. \c{S}eny\"uz\altaffilmark{1}}
\affil{Department of Physics, Faculty of Arts and Sciences, \c{C}anakkale Onsekiz Mart University, TR-17100 \c{C}anakkale, Turkey}


\altaffiltext{1}{Astrophysics Research Center and Ulup{\i}nar Observatory, \c{C}anakkale Onsekiz Mart University, TR-17100, \c{C}anakkale, Turkey}


\begin{abstract}
The young binary system HD 350731 is a noteworthy laboratory for studying early-type binaries with similar components. We present here the analysis of differential multi-color photometric and spectroscopic observations for the double-lined detached system. Accurate absolute parameters were determined from the simultaneous solution of light and radial velocity curves for the first time. HD 350731 consists of two B8V-type components having masses and radii respectively of $M_{1}=2.91\pm0.13$\,M$_{\odot}$, $M_{2}=2.80\pm0.14$\, M$_{\odot}$, $R_{1}= 2.11 \pm0.05$\,R$_{\odot}$ and $R_{2}=2.07\pm0.05$\,R$_{\odot}$. The effective temperatures were determined based on analysis of disentangled spectra of the components and derived to be $12000\pm250$ K and $11830\pm300$ K for the primary and secondary components, respectively. The measured projected rotational velocities, 69.2$\pm$1.5 km s$^{-1}$ for primary and 70.1$\pm$1.7 km s$^{-1}$ for secondary, were found closer to the pseudo-synchronous velocities of the components. Comparison with evolutionary models suggests an age of 120$\pm35$ Myr. Kinematic analysis of the unevolved binary system HD 350731 revealed that it belongs to the young thin-disc population of the Galaxy.
\end{abstract}





\section{Introduction}

Precise knowledge of the absolute parameters of stars (mass, radii, etc.) is a means to better understand the structure and evolution of galaxies. Eclipsing binary stars are the most important objects for determining these basic parameters of stars. In order to derive the mass, radius, temperature and other physical parameters of eclipsing binary stars, spectroscopic, photometric -and interferometric data are required. In particular, the quality of the observational data plays crucial role in determining the absolute parameters of the component stars precisely. Among the eclipsing binary stars, detached double-lined spectroscopic binaries are the best sources for acquiring more accurately the main physical properties of stars (e.g. Southworth 2013, Lacy et al. 2015). Well measured detached eclipsing binaries have been recently listed and studied based on distributions of their absolute parameters by Torres et al. (2010) and Eker et al. (2014).

\object{HD 350731} (BD+20 4323, GSC 01624-00493, $V=9^{m}$.60) was identified as an eclipsing binary system with eccentric orbit by Otero et al. (2004). In the AGK3 catalogue, the spectral type of the system is given as A0 by Heckmann (1975) based on the Henry Draper (HD) catalogue (Pickering and Cannon 1918-1924). However, Nesterov et al. (1995) suggested it was B9 using the extension charts of HD catalogue. The first photometric study, based on $BVR_{c}I_{c}$ light curves, was conducted by Kleidis et al. (2008) in which, in addition to preliminary photometric analysis, it was suggested that HD 350731 indicates apsidal motion.

In this study, the first detailed photometric and spectroscopic analysis was carried out, based on newly-obtained data. After descriptions of the spectroscopic and photometric observations, Section 3 presents spectral analysis, which contains radial velocity measurements of the components, preliminary orbit solution, spectroscopic light ratio, spectral disentangling and model atmosphere application. Analysis of multi-color light and radial velocity curves is given in Section 4, followed by apsidal motion analysis. The absolute parameters and kinematic properties of the system are detailed in Sect. 6. Finally, we conclude the study with the results and discussions.

\section{Observations}

New multi-color photometric observations of \object{HD 350731} were performed at \c{C}anakkale Onsekiz Mart University Ulup{\i}nar Observatory, Turkey over 11 nights in August and September 2012. Observations were carried out with a 60 cm Cassegrain telescope equipped with SBIG STL1001E CCD camera. One secondary minimum was also observed with a 122 cm Cassegrain-Nasmyth telescope incorporating Apogee Alta U42 CCD camera. The data were collected using Bessell $B$, $V$ and $R_{c}$ filters. HD 350730 (A0) and HD 350727 (F5) were used as comparison and check star, respectively. CCD frames were reduced in the standard way: bias and dark frames were subtracted from the frames and then corrected for flat-fielding. Such reduced images were used to extract the differential magnitudes of HD 350731. C-Munipack\footnote[1]{http://c-munipack.sourceforge.net/} code was used for these processes. Standard deviation of variations between the observed comparison and check stars was determined to be about 0$^{m}$.01 for all passbands.

Spectroscopic observations of \object{HD 350731} were made using the Cassegrain spectrograph installed on the 1.85 m Plaskett telescope at Dominion Astrophysical Observatory (DAO), British Columbia, Canada. The spectrograph has a spectral resolving power of about $R=9000$. A back-illuminated SITe CCD of 1752$\times$532 pixels (size 15 $\mu$m) was used to record spectra spanning from about 4370 to 4630 {\AA}. During the observations, twenty one spectra for \object{HD 350731} and several spectra for 21 Peg were taken. The spectra of the reference star 21 Peg were used to measure the radial velocities of the components. All spectral data were acquired between 6 August and 6 September 2012. Spectroscopic data reduction was handled using the appropriate tasks of IRAF\footnote[2]{http://iraf.noao.edu/} package with the following steps: background subtraction, division by flat field spectrum, wavelength calibration using Fe-Ar lamp, and normalization to the continuum.

\section{Spectroscopic Analysis}

\subsection{Radial Velocities and Preliminary Orbital Solution}
In order to determine the orbital parameters of \object{HD 350731}, the radial velocities (RVs) of the components must be measured. Taking into account the spectral type of the system, given as B9 by Nesterov et al. (1995), 21 Peg (B9.5V, $V_r=-0.2$ km s$^{-1}$) was chosen as the RV standard star for both components. The cross-correlation technique (CCT) was used for determination of RVs of the components. Simkin (1974) reported an useful description of CCT which are widespread, especially for RV measurements of binary stars (e.g. Hill 1993, Gunn et al. 1996, Frasca et al. 2002, Soydugan et al. 2007). The CCT was applied by FXCOR routine in the IRAF package which was developed based on the standard Tonry \& Davis (1979) algorithm. The weights of RVs and their standard errors were calculated according to the usual formulas given by Topping (1972) and Tonry \& Davis (1979). The measured RVs of both components are given in Table 1 together with their standard errors, which are between 4-9 km s$^{-1}$ for the components.

\begin{table}
%\scriptsize
 \begin{center}
 \small
  \caption{Radial velocity measurements of components of HD 350731.}
  \begin{tabular}{@{}llll@{}}
  \hline\hline
  HJD       & Orbital   & \multicolumn{1}{c} {$V_{1}$}       & \multicolumn{1}{c} {$V_{2}$}\\
  24 50000+ & Phase     & \multicolumn{1}{c} {(km s$^{-1}$)} & \multicolumn{1}{c} {(km s$^{-1}$)}\\
  \hline
6173.8976 & 0.0773 &    -72.9  $\pm$ 5.8 &  49.1    $\pm$ 8.1 \\
6152.7108 & 0.1200 &    -112.8 $\pm$ 5.9 &  93.5    $\pm$ 7.1 \\
6152.7482 & 0.1429 &    -126.7 $\pm$ 6.5 &  111.1   $\pm$ 7.8 \\
6147.8861 & 0.1694 &    -136.2 $\pm$ 6.2 &  128.1   $\pm$ 7.0 \\
6152.8407 & 0.1995 &    -146.9 $\pm$ 6.8 &  134.8   $\pm$ 8.1 \\
6175.8132 & 0.2488 &    -155.6 $\pm$ 6.7 &  137.6   $\pm$ 8.4 \\
6152.9240 & 0.2505 &    -152.9 $\pm$ 6.6 &  137.7   $\pm$ 8.2 \\
6175.8962 & 0.2996 &    -161.0 $\pm$ 6.8 &  132.6   $\pm$ 8.1 \\
6180.8738 & 0.3437 &    -139.1 $\pm$ 6.2 &  129.4   $\pm$ 6.2 \\
6149.8223 & 0.3536 &    -136.5 $\pm$ 6.4 &  116.6   $\pm$ 8.9 \\
6149.9104 & 0.4074 &    -97.0  $\pm$ 6.6 &   89.7   $\pm$ 7.1 \\
6172.8880 & 0.4598 &    -68.1  $\pm$ 6.5 &   54.1   $\pm$ 7.4 \\
6177.8676 & 0.5052 &    -10.9  $\pm$ 4.0 &  \multicolumn{1}{c} {--}\\
6146.9266 & 0.5826 &    45.7   $\pm$ 5.2 &   -62.3  $\pm$ 7.7 \\
6174.7848 & 0.6199 &    64.2   $\pm$ 6.8 &   -94.4  $\pm$ 7.2 \\
6151.9360 & 0.6462 &    99.9   $\pm$ 6.5 &   -124.1 $\pm$ 7.1 \\
6148.7783 & 0.7151 &    142.2  $\pm$ 7.7 &   -172.6 $\pm$ 7.8 \\
6148.8464 & 0.7567 &    162.4  $\pm$ 7.1 &   -181.1 $\pm$ 7.4 \\
6148.9094 & 0.7952 &    155.1  $\pm$ 6.8 &   -182.8 $\pm$ 8.5 \\
6171.8986 & 0.8548 &    136.5  $\pm$ 8.4 &   -158.5 $\pm$ 8.2 \\
6176.8712 & 0.8958 &    98.8   $\pm$ 8.8 &   -127.8 $\pm$ 6.8 \\
\hline
\end{tabular}
\end{center}
\end{table}

As can be seen in the light curves of the system, the secondary minimum is not located at the 0.5 orbital phase and indicates a displacement comparing primary minimum. Therefore, most probably \object{HD 350731} has an eccentric orbit and this must be taken into account for the orbital solution. Using the RVs listed in Table 1 and adopted ephemeris from Kleidis et al. (2008), we obtained the orbital parameters of the system listed in Table 2 in order to have input data for KOREL application and also the simultaneous analysis of light and radial velocity curves. The preliminary orbit parameters together with their uncertainties are given in Table 2.

\begin{table}
  \caption{Orbital elements of HD 350731.}
  \label{tab:orbit}
  \begin{center}
  \small
           \begin{tabular}[h]{lll}
      \hline\hline
          Parameter    & &    Value                     \\
      \hline
      \emph{T$_{0}$} (HJD)          & & 2454631.4603\footnotemark[1]  \\
      \emph{P$_{orb}$ }(day)        & & 1.635135\footnotemark[1]     \\
      \emph{V$_{\gamma}$ }(km s$^{-1}$) & & -10.4 $\pm$ 0.7                \\
      \emph{K$_{1}$ } (km s$^{-1}$) & & 157.2 $\pm$ 1.3               \\
      \emph{K$_{2}$ } (km s$^{-1}$) & & 162.7 $\pm$ 1.3              \\
      \emph{e}                      &  & 0.077 $\pm$ 0.007      \\
      \emph{$\omega$} (degree)             &  & 23.5 $\pm$ 2.3      \\
      \emph{a$_{1}\sin i$} (10$^{6}$\,km) &  & 3.52 $\pm$ 0.04         \\
      \emph{a$_{2}\sin i$} (10$^{6}$\,km) &  & 3.64 $\pm$ 0.04         \\
      \emph{M$_{1}\sin^{3}$\,$i$} (M${_{\sun}}$) & &  2.79 $\pm$ 0.05 \\
      \emph{M$_{2}\sin^{3}$\,$i$} (M${_{\sun}}$) & &  2.70 $\pm$ 0.05 \\
      \emph{q (=M$_{2}$/M$_{1}$)}   &  & 0.966 $\pm$ 0.015      \\
         \hline
     \end{tabular}\\
      \footnotemark[1]{Kleidis et al. (2008)}\\
      \end{center}
      \end{table}

\subsection{Spectroscopic Light Ratio}

Spectroscopic line ratios ensure us constraints on the luminosity ratios of the components determined from the light curve solutions at similar wavelengths, as reported in the study of Petrie (1939) and applied in many studies (e.g. Andersen et al. 1983, Southworth \& Clausen 2007, Garcia et al. 2014). This also helps us to reduce the degeneracy, which occurs in the light curve solutions of eclipsing binaries with similar components for the radii of the components, if the binary are partially eclipsing system.
In this study, we have derived spectroscopic light ratio for the system using the spectral line of MgII at 4481 {\AA}. Equivalent widths (\emph{EW}) of MgII lines for the both components were measured using SPLOT task in the IRAF package on seven spectra taken at the phases out of the eclipses. The weighted mean ratio was found to be \emph{EW$_{2}/EW_{1}$} = 0.953$\pm$0.042, which can be accepted to be the light ratio of the components around 4481 {\AA}. Since the components of HD 350731 are almost equal and have very similar spectral properties, the spectral correction and wavelength dependence for light ratio are negligible.

\subsection{Spectral Disentangling}

In order to derive the atmospheric parameters of the components of \object{HD 350731}, individual spectrum for each component is required. The method of spectral disentangling of the composite spectra was invented and described by Simon \& Sturm (1994). A well discussion and assessment on this approach was given by Pavlovski \& Hensberge (2010) and Pavlovski \& Southworth (2012). The disentangled method was also preferred in this study to decompose the observed spectra of the system to its components, as applied in some studies (e.g. Groenewegen et al. 2007, Torres et al. 2011, Lehmann et al. 2013, Harmanec et al. 2014). The KOREL code presented by Hadrava (1995) was used for spectral disentangling. Fortran code KOREL was developed based on Fourier analysis technique to disentangle component spectra in binary or multiple stellar systems.

Nineteen spectra of the system obtained at DAO were used for the disentangling process. For the application, wavelength region 4375-4575{\AA}, which includes Mg II and He I spectral lines, was used. The orbit properties except systemic velocity of the the system $(V_{\gamma})$ could be determined with KOREL. However, the radial velocities of the components derived from KOREL are not independent as noted by Lehmann et al. (2013). Therefore, we used the orbital properties listed in Table 2 as input data for KOREL application. Finally, we derived the radial velocity amplitudes ($K_{1}$ and $K_{2}$) from two different methods, which are compatible within better than 2 km s$^{-1}$. The fractional light contribution of the components determined from spectroscopic analysis was used for the application. At the end of the process, separate spectra of the primary and secondary components were obtained. Some observed composite spectra of the system at different orbital phases, together with the model spectra and decomposed spectrum of the components calculated by KOREL in the wavelength range 4450-4500\,{\AA}, are shown in Figure 1.

\begin{figure}
\begin{center}
\includegraphics[width=105mm,height=125mm]{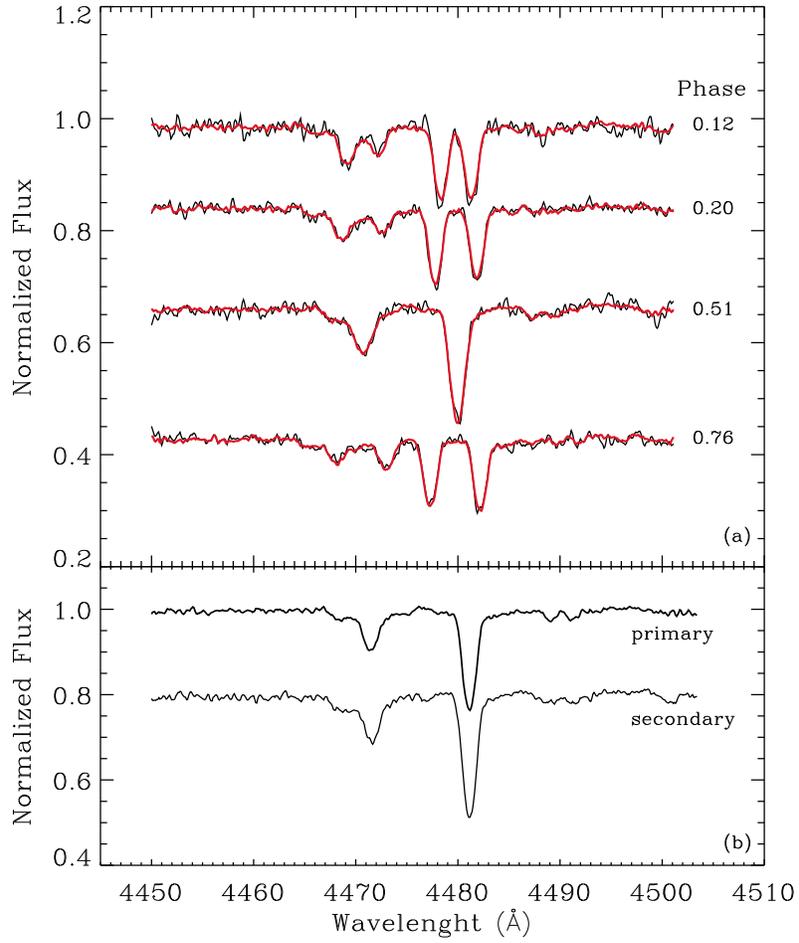}
\caption{Fits calculated by KOREL (thick lines) to observed spectra (thin lines) at different orbital phases (a), and disentangled spectra of primary and secondary components (b) in wavelength range 4450-4500{\AA}.}
\end{center}
\end{figure}

\subsection{Model Atmosphere Application}

The physical parameters of the components of \object{HD 350731} can be derived from analysis of the radial velocities of the components and light curves. On the other hand, if one can obtain decomposed spectra of the components, their atmospheric parameters can be found by line-profile fitting.
The advantages of spectral disentangling method for the spectroscopic analysis of the eclipsing binaries' components are presented in the studies of Hensberge et al. (2000) and Pavlovski \& Hensberge (2005). Therefore, we obtained the individual spectrum of the components of HD 350731 considering these previous studies to make spectral analysis. The SME (Spectroscopy Made Easy) code, which enabled us to determine the basic atmospheric parameters of the components by matching the computed spectrum to the observed one was used for the analysis. SME was developed by Valenti \& Piskunov (1996) and includes several model atmospheres to compute synthetic spectra. The code uses Levenberg-Marquardt algorithm to fit an observing spectrum with a synthetic one. The code has been applied in many studies to determine the atmospheric properties of stars (e.g. Torres et al. 2012, Soydugan et al. 2013). For the application, Kurucz (1993) model atmospheres were used and the atomic data for the spectral lines were taken from the Vienna Atomic Line Database (Piskunov et al. 1995; Kupka et al. 1999).

The disentangled spectrum of both components in the wavelength region 4375-4575{\AA} included Mg II at 4481 \AA\ and He I at 4471 \AA\ spectral lines. These data were used to compute of the model atmospheres. The microturbulent velocities for the components were adopted to be 3 km s$^{-1}$, which is an appropriate value for late-type B stars (e.g. Adelman 1996, Usenko et al. 2000). The surface gravities of both components were fixed at the dynamical values determined from solution light and radial velocity curves (see Table 8). After preparing SME code to fit the spectrum of the components of HD 350731, atmospheric parameters, namely T$_{eff}$ and \emph{$v\sin i$} adjusted during the analysis have been determined assuming solar abundance taken from Asplund et al. (2009). The resulted parameters and their uncertainties estimated using the $\Delta \chi^{2}_{min}=1$ method as described by Lampton et al. (1976) are listed in Table 3 . In Fig. 2, the decomposed spectra of the components and the synthetic spectra calculated by the best model parameters in Table 3 are compared. The figure also shows the observed composite spectrum of \object{HD 350731} together with the computed spectrum, which was calculated by the model atmosphere parameters in Table 3, taking into account light contributions at the orbital phase of 0.2. The synthetic composite binary spectrum in Fig. 2 was calculated by using BinMag IDL visualization code (developed by O. Kochukov). As seen in Fig. 2, the synthetic spectra agree well with the composite binary spectrum and individual spectra of the components.

\begin{table}
  \caption{Model atmosphere parameters of the components of HD 350731.}
  \label{tab:model_atm}
  \begin{center}
  \small
           \begin{tabular}[h]{llll}
      \hline\hline
                       & &     Primary       & Secondary   \\
          Parameter    & &     Value         & Value           \\
      \hline
      $T_{eff}$ (K)             & & 12000 $\pm$ 250  & 11830 $\pm$ 300 \\
      $\log g$ (cgs)            & & 4.25\footnotemark[1]   & 4.25\footnotemark[1]            \\
      $v\sin i$ (km s$^{-1}$)   & & 69.2 $\pm$ 1.5   & 70.1 $\pm$ 1.7 \\
            \hline
           \hline
     \end{tabular}\\
     \footnotemark[1]{Dynamical values adopted from analysis of light and radial velocity curves.}\\
      \end{center}
      \end{table}

\begin{figure}
\begin{center}
\includegraphics[width=125mm,height=80mm]{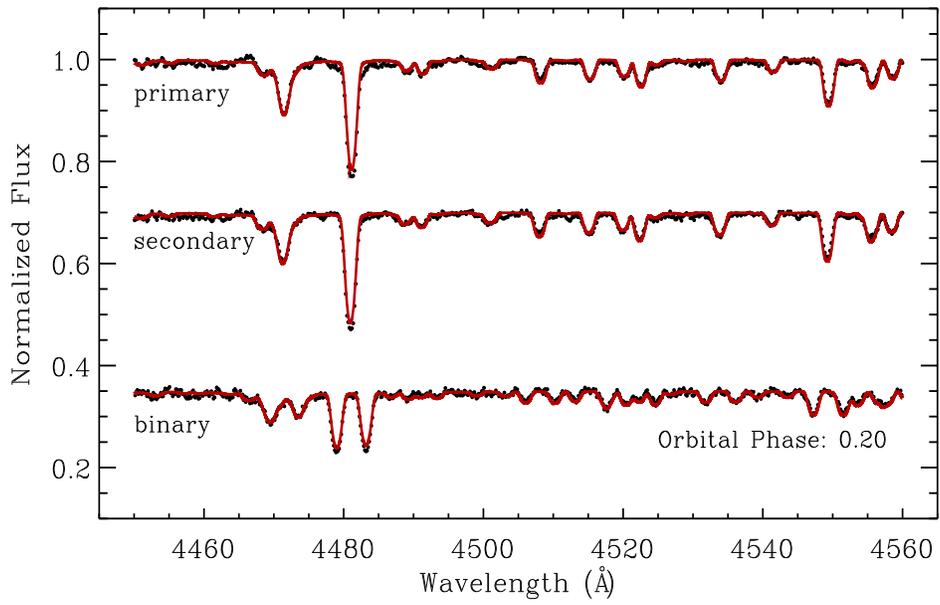}
\caption{Disentangled spectra of components with best model atmosphere fits. Comparison between observed composite binary spectrum and computed spectrum calculated with parameters in Table 3.}
\end{center}
\end{figure}

\clearpage

\section{Analysis of Light and Radial Velocity Curves}

The $BVR_{c}$ light curves and radial velocities of the components of \object{HD 350731} have been modeled using the version v34 of JKTEBOP code (Southworth et al. 2004, Southworth et al. 2005a, Southworth 2013). The code is based on the EBOP program, which was developed by P. Etzel (Etzel 1981, Popper \& Etzel 1981). It was written in FORTRAN77 by J. Southworth\footnote[3]{http://www.astro.keele.ac.uk/~jkt/codes/} and uses Levenberg-Marquardt algorithm to reach the best model. We preferred JKTEBOP code since it was stable, and very fast and includes various error-estimate algorithms. This code is also very useful for well-detached eclipsing binaries as tested in several studies (e.g. Ratajczak et al. 2010, Debosscher et al. 2013, Lehmann et al. 2013). 

\object{HD 350731} is a partially eclipsing binary with almost identical components. In that case, the degeneracy may occur in determination of the radii of the components. Therefore, we used spectroscopic line ratio as given in the Section 3.2 to constrain the range of the ratio of the radii ($k$). As the first step, the luminosity ratio was adopted at 0.953 as determined from the spectroscopy during the analysis of $B$ light curve since MgII line at 4481 {\AA} was used to derive spectroscopic light ratio. Thus, the $k$ value was found to be $0.993\pm0.030$. In order to check the $k$ value and also examine the distributions of $k$ values against to surface brightness ratio of the components ($J_{2}/J_{1}$), these two parameters were scanned corresponding $\chi^{2}$ values of each solution. Then, $\chi^{2}$ values of the solutions were mapped into contours which indicate the uncertainties in Fig. 3. As shown in the figure, the $k$ value of 0.993 can be seen around the lowest $\chi^{2}$ value. After that, the resulted $k=0.993$ was adopted during the analysis of $V$ and $R_{c}$ light curves. The third light contribution (\emph{l$_{3}$}) was assumed to be 0.0 after several iterations since it did not vary significantly. We have used the linear limb darkening coefficients as adopted parameters for the components corresponding to photometric filters used since the quadratic and logarithmic laws did not give better fits. The adjustable parameters are central surface brightness ratio of the components ($J_{2}/J_{1}$), sum of the fractional radii ($r_{1}+r_{2}$), ratio of fractional radii ($k$) for only $B$ filter, orbital inclination ($i$), eccentricity ($e\,\sin\omega$, $e\,\cos\omega$), phase shift, radial velocity amplitudes of the components ($K_{1}$ and $K_{2}$) and the systemic velocity ($V\gamma$).

\begin{figure}
\begin{center}
\includegraphics[width=120mm,height=150mm]{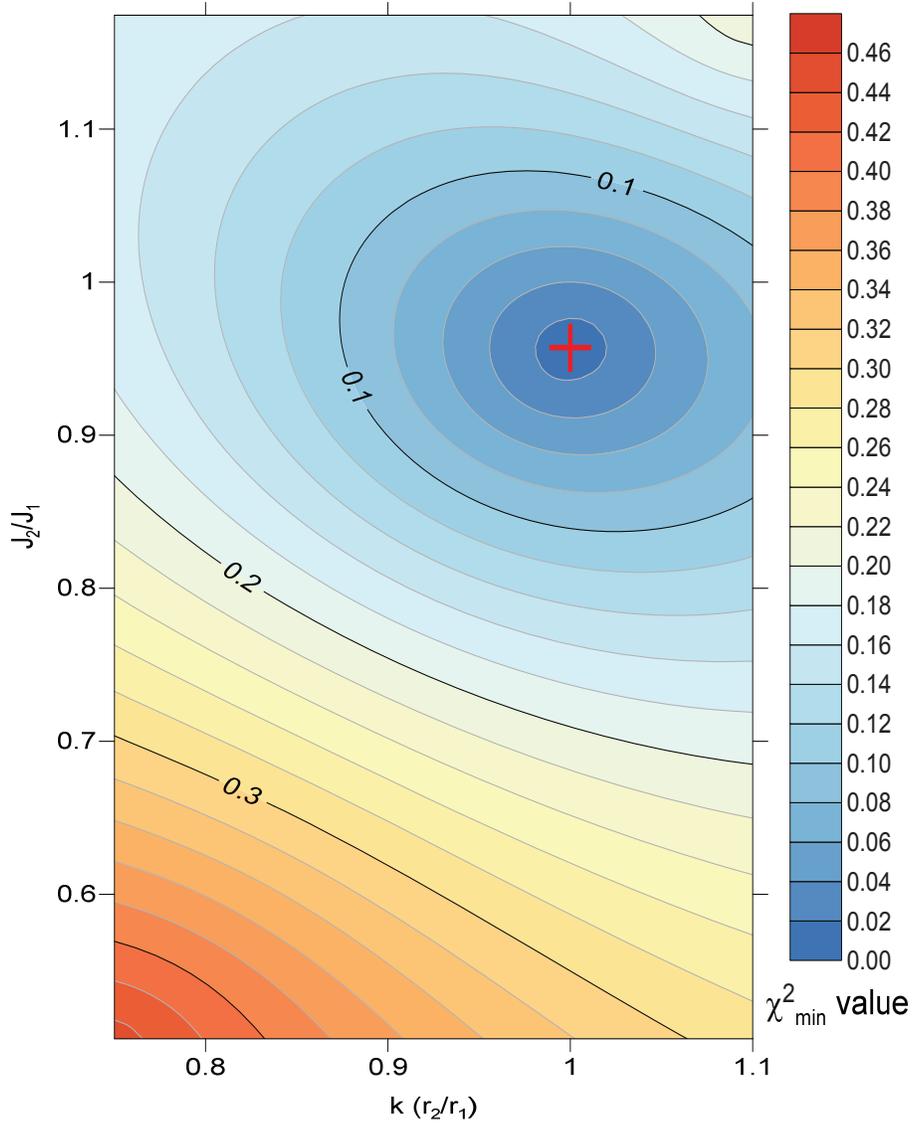}
\caption{Contour map indicating the distribution of the results from $B$ light curve analysis for the correlated parameters of surface brightness ratio ($J_{2}/J_{1}$) and ratio of fractional radii ($k$) of the components. The position of the plus symbol in the figure represents minimum $\chi^{2}$ value.}
\end{center}
\end{figure}

\clearpage

In order to determine uncertainties for the solutions, we used task 9 of JKTEBOP code and calculated 1$\sigma$ errors with Monte Carlo algorithm. The resulted parameters are presented in Table 4 together with their uncertainties for $B, V$ and $R_{c}$ filters. The adopted photometric parameters, which are the weighted mean of the solutions in each filter, are listed in Table 5. In this table, the final orbital parameters and also their uncertainties calculated by JKTEBOP are also given. A comparison between observed and computed light curves is presented in Fig. 4 together with the residuals of the observational data from the best fits, while the RVs of the components together with the best fits are shown in Fig. 5.

\begin{table}
  \begin{center}
  \small
  \caption{Parameters of HD 350731 obtained from analysis of multi-color light curves and radial velocities of components.}\label{tbl-3}
    \begin{tabular}{llll}
  \hline\hline
 Parameters                  & $B$        &    $V$    &   $R_c$     \\
 \hline
 $r_{1}+r_{2}$               & 0.399$\pm$0.003 & 0.400$\pm$0.003  & 0.404$\pm$0.004 \\
 $k$                         & 0.993$\pm$0.030 & 0.993\footnotemark[1]  & 0.993\footnotemark[1] \\
 $r_{1}$                      & 0.2000$\pm$0.0031 & 0.2009$\pm$0.0014  & 0.2028$\pm$0.0018    \\
 $r_{2}$                      & 0.1986$\pm$0.0031 & 0.1995$\pm$0.0014  & 0.2014$\pm$0.0018    \\
 \textit{i}\,(\degr)            & 81.83$\pm$0.09  & 81.75$\pm$0.10  & 81.40$\pm$0.11   \\
 $J_{2}/J_{1}$                & 0.968$\pm$0.010  & 0.977$\pm$0.011  & 0.976$\pm$0.015 \\
 $e\,\sin\omega$               & 0.0349$\pm$0.0060  & 0.0310$\pm$0.0020  & 0.0330$\pm$0.0078 \\
 $e\,\cos\omega$               & 0.0719$\pm$0.0004  & 0.0720$\pm$0.0008  & 0.0717$\pm$0.0006 \\
 \hline
\end{tabular}\\
 \footnotemark[1]{Fixed during the analysis.}\\
\end{center}
\end{table}

\begin{figure}
\epsscale{.70} \plotone{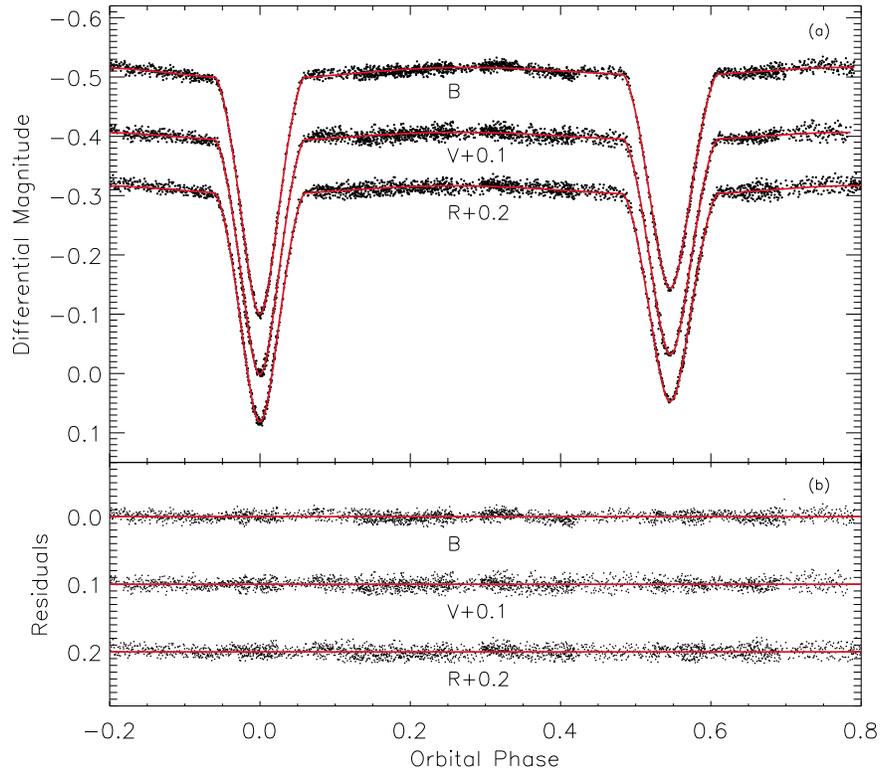} \caption{Observed and theoretical light curves of HD 350731 in $BVR_c$ filters (a) and the residuals from the best fits (b).}
\end{figure}

\begin{table}
  \begin{center}
  \small
  \caption{Final values of photometric and spectroscopic parameters for HD 350731.}\label{tbl-3}
    \begin{tabular}{llll}
  \hline\hline
 Parameters             & Value        \\
 \hline
 $r_{1}$                & 0.2014$\pm$0.0010 \\
 $r_{2}$                & 0.2000$\pm$0.0010 \\
 \textit{i}\,(\degr)    & 81.703$\pm$0.061   \\
 $L_{2}/L_{1}\,\, (B)$      & 0.952$\pm$0.012 \\
 $L_{2}/L_{1}\,\, (V)$      & 0.961$\pm$0.011 \\
 $L_{2}/L_{1}\,\, (R_c)$      & 0.961$\pm$0.014 \\
 $e$                    & 0.0792$\pm$0.0017   \\
 $\omega$\,(\degr)           & 24.63$\pm$2.42  \\
 $K_{1}$ (km s$^{-1}$)  & 157.44$\pm$2.77 \\
 $K_{2}$ (km s$^{-1}$)               & 163.14$\pm$2.22 \\
 $V_{\gamma}$ (km s$^{-1}$)          & -10.29$\pm$1.28 \\
 \hline
\end{tabular}
\end{center}
\end{table}

\begin{figure}
\begin{center}
\includegraphics[width=110mm,height=90mm]{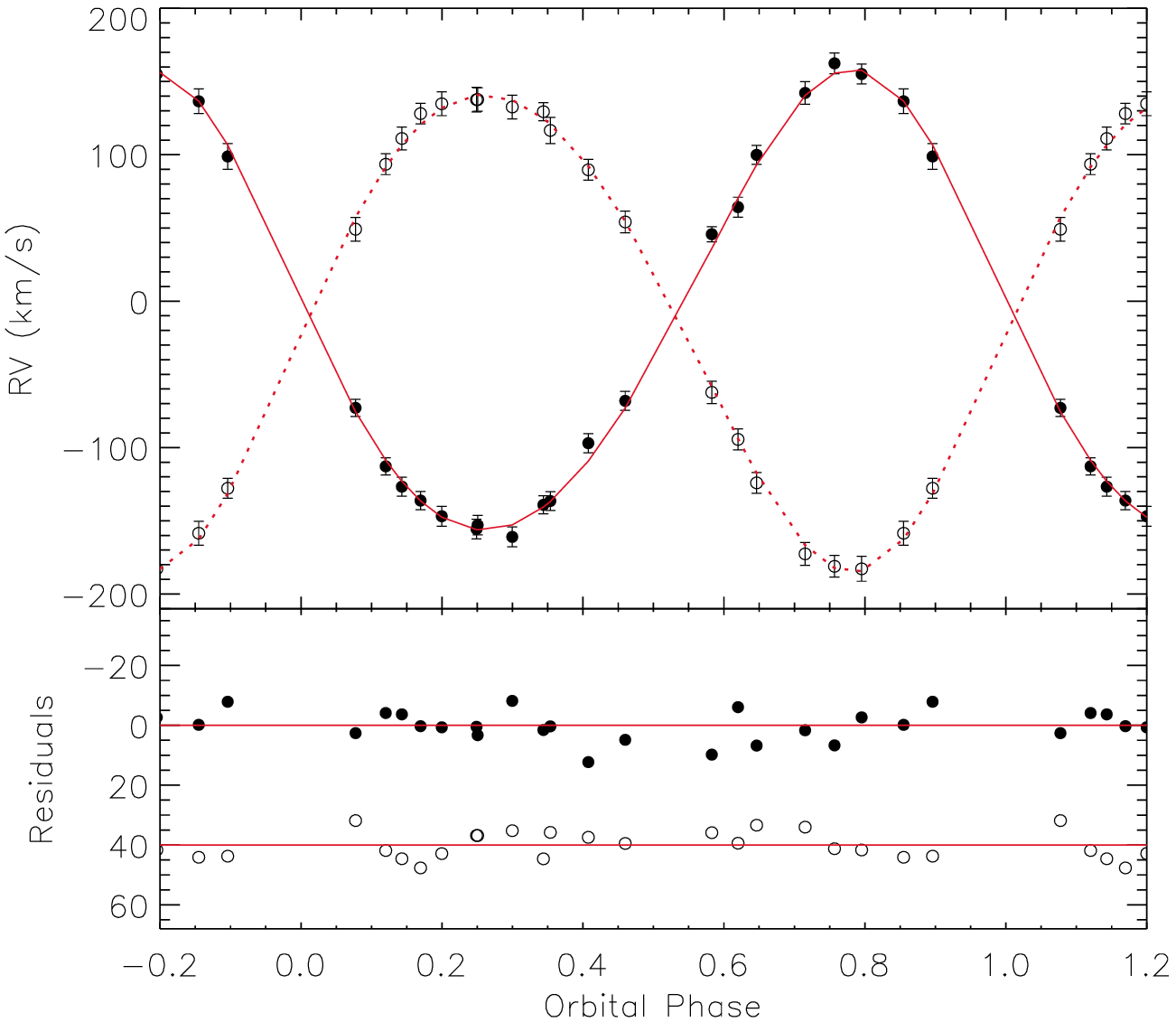}
\caption{RV curves of components of HD 350731 plotted
versus orbital phase. Filled and open circles
represent RVs of primary and secondary components, respectively.
Solid line represents orbital solution for primary
component and dashed line for secondary component. The residuals are indicated below.}
\end{center}
\label{fig:RV_HD 350731}
\end{figure}

\section{Apsidal Motion Analysis}

The apsidal motion of \object{HD 350731} was indicated for the first time by Kleidis et al. (2008). In order to establish a preliminary estimation of the apsidal motion elements based on the commonly-used method of $O-C$ data analysis, we collected published minima times and determined two primary and two secondary minima times from our observations, as listed in Table 6. In total, 15 minima times were achieved.

The analysis was made using code written by Zasche et al. (2009), which was developed based on the mathematical methods reported by Gim\'{e}nez \& Garcia-Pelayo (1983). For the analysis, the orbital inclination and eccentricity were adopted as $i=81^{\circ}.70$ from the light curve analysis and $e$=0.079 from the orbital analysis. The resulting preliminary apsidal motion parameters are given in Table 7, while the $O-C$ diagram with theoretical fits is shown in Fig. 2. The changing rate in longitude of the periastron is $\dot{\omega}=0.0170\pm0.0082$ degree cycle$^{-1}$, which corresponds to an apsidal motion period of \emph{U}=92 yr. As seen in Table 7, the errors of the parameters are high and the parameters are not sensitive since 15 minima times, which cover only 9 yr, were used for the analysis. However, the results can be accepted for a preliminary estimation of the apsidal motion properties and improved by adding new times of minima.

\begin{table}
%\scriptsize
 \begin{center}
 \small
  \caption{List of minima times of HD 350731.}
  \begin{tabular}{@{}lllll@{}}
  \hline\hline
  HJD       & Error & Filter       &   Epoch  &     Ref. \\
  24 50000+ &       &              &          &          \\
  \hline
3594.4279 & --     & $I$           & -646.5  &    1\\
3612.4138 & 0.0001 & $R_{c}$       & -635.5  &    2\\
3653.2919 & 0.0001 & $R_{c}$       & -610.5  &    2\\
4002.3133 & 0.0001 & $R$           & -397.0  &    1\\
4299.9065 & 0.0002 & $V$           & -215.0  &    3\\
4340.7838 & 0.0002 & $V$           & -190.0  &    3\\
4345.6896 & 0.0001 & $V$           & -187.0  &    3\\
4628.5684 & 0.0002 &$BVR_{c}I_{c}$ & -14.0   &    3\\
4642.5482 & 0.0002 &$BVR_{c}I_{c}$ & -5.5    &    3\\
4651.4605 & 0.0004 &$BVR_{c}I_{c}$ & 0.0     &    3\\
4665.4390 & 0.0002 &$BVR_{c}I_{c}$ & 8.5     &    3\\
6175.4058 & 0.0001 &$BVR_{c}$      & 932.0   &   this study\\
6176.2986 & 0.0002 &$BVR_{c}$      & 932.5   &   this study\\
6180.3108 & 0.0003 &$BVR_{c}$      & 935.0   &   this study\\
6892.4851 & 0.0001 &$VR_{c}$       &1370.5   &   this study\\
\hline
\end{tabular}\\
 \footnotemark[1]{Br\'{a}t et al. (2007),}
 \footnotemark[2]{Zejda et al. (2006),}
 \footnotemark[3]{Kleides et al. (2008).}
\end{center}
\end{table}

\begin{table}
  \caption{Apsidal motion elements of HD 350731.}
  \label{tab:apsidal}
  \begin{center}
  \small
           \begin{tabular}[h]{lll}
      \hline\hline
          Parameter    & &    Value                     \\
      \hline
      \emph{T$_{0}$} (HJD)         &           & 2454651.5008 $\pm$ 0.0102  \\
      \emph{P$_{s}$ }(day)         &           & 1.63511 $\pm$ 0.00001   \\
      \emph{e}                     &           & 0.079$^{a}$      \\
      \emph{$\omega$} (degree)     &           & 3.4 $\pm$ 4.3      \\
      \emph{d$\omega$/dt} (degree cycle$^{-1}$)&& 0.0170 $\pm$ 0.0082      \\
      \emph{U}(yr)                 &           & 92 $\pm$ 35        \\
      \emph{P$_{a}$ }(day)         &           & 1.63519 $\pm$ 0.00002\\
      \hline
     \end{tabular}\\
     \footnotesize $^{a}$: Adopted from orbital analysis.
      \end{center}
      \end{table}

\begin{figure}
\begin{center}
\includegraphics[width=140mm,height=80mm]{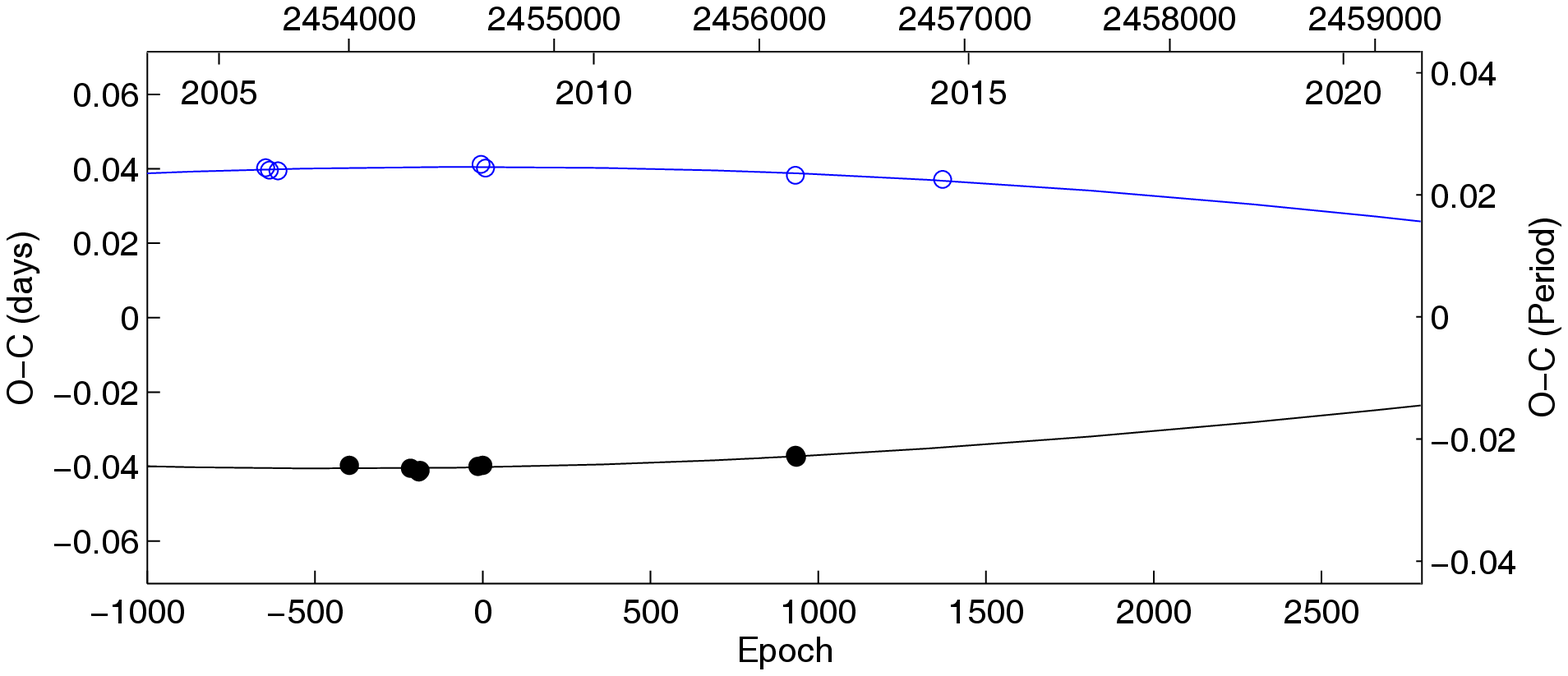}
\caption{$O-C$ diagram for HD 350731 obtained with parameters given in Table 7.
Primary and secondary minima times are indicated by filled and open circles, respectively.}
\end{center}
\label{fig:O-C}
\end{figure}

\section{Absolute Properties and Kinematic Behavior}
Analysis of the light curve of \object{HD 350731} and the radial velocities of its components led to acquisition of its accurate absolute dimensions and physical properties for the first time. In order to calculate absolute parameters and their uncertainties except effective temperatures, which were determined from spectral analysis, we used the JKTABSDIM code, which is developed by Southworth et al. (2005b). In this code, the uncertainties are calculated with very robustly and a complete error budget is found for every output parameter. Derived fundamental parameters are listed in Table 8. The physical constants used for the calculations are given by Southworth (2011).

The accuracy of the masses is better than 5\%, while radii values of the components are good to about 2.5\%. We calculated $E(B-V)$ color excess for \object{HD 350731} from  the dust maps of Schlafly \& Finkbeiner (2011). Because the system is relatively close to the sun, the color excess value from Schlafly \& Finkbeiner (2011) needed to be reduced according to the distance. The $J$ band absolute magnitude of the system was calculated using the color-luminosity relation of Bilir et al. (2008). Then, the calculated color excess $E(B-V)=0.157$ mag was reduced using the equation of Bahcall \& Soneria (1980). The color excess value is consistent with the position of \object{HD 350731} in the Milky Way. The interstellar absorption value (\emph{A$_{v}$}) has been calculated to be 0.49 mag from the commonly-accepted formula \emph{A$_{v}=3.1\times E(B-V)$}. The distance of the system was found to be 703$\pm$34 pc based on apparent system magnitude, light ratio of the system's components, and interstellar extinction values. The temperatures of the components determined from spectroscopic analysis are consistent with the spectral types of B8V+B8V, according to calibrations given by Sung et al. (2013).

The components of the system have similar properties since the temperature difference between the components was found to be $\Delta$\emph{T}=170 K which is smaller than 1$\sigma$ uncertainties for the effective temperatures of both components (see Table 3). Furthermore, the masses and radii values were very close, as seen in Table 8. From the photometric analysis, the system was found to be detached and the Roche lobe filling ratios were calculated to be 63\% and 64\% for the primary and secondary components, respectively.

\begin{table}
  \caption{Astrophysical properties of HD 350731.}
  \label{tab:par}
  \begin{center}
             \begin{tabular}[h]{lcc}
  \hline\hline
     Parameter                                 &    Primary    & Secondary  \\
  \hline
     Mass (M$_{\odot}$)                         & 2.91$\pm$0.13 &  2.80$\pm$0.14 \\
     Radius (R$_{\odot}$)                       & 2.11$\pm$0.05 &  2.07$\pm$0.05  \\
     Temperature (K)                            & 12000$\pm$250 & 11830$\pm$300  \\
     $\log L$ (L$_{\odot}$)                     & 1.92$\pm$0.04 & 1.88$\pm$0.05 \\
     $\log g$ (cgs)                             & 4.25$\pm$0.02 & 4.25$\pm$0.02 \\
     Orbital period (day)                       & \multicolumn{2}{c}{1.63511$\pm$1$\times$10$^{-5}$} \\
     Orbital separation (R$_{\odot}$)           & \multicolumn{2}{c}{10.43$\pm$0.02}          \\
     Mass ratio                                 & \multicolumn{2}{c}{0.965$\pm$0.004} \\
     Systemic velocity (km\,s$^{-1}$)           & \multicolumn{2}{c}{-10.1$\pm$0.4}  \\
     Distance (pc)                              & \multicolumn{2}{c}{703$\pm$34}   \\
     \emph{V} (mag)                             & \multicolumn{2}{c}{9.60$^{a}$}  \\
     \emph{M$_{Bol}$} (mag)                     & -0.05$\pm$0.11 & 0.05$\pm$0.12     \\
     \emph{BC} (mag)                            & -0.66$^{b}$        &    -0.62$^{b}$   \\
     \emph{M$_{V}$} (mag)                       & 0.61$\pm$0.11& 0.67$\pm$0.12           \\
     Measured $v\sin i$ (km\,s$^{-1}$)          & 69.2$\pm$1.5  & 70.1$\pm$1.7           \\
     Pseudo-synchronous $v\sin i$ (km\,s$^{-1}$) & 67.5$\pm$0.3 & 67.0$\pm$0.3           \\
     Synchronous $v\sin i$ (km\,s$^{-1}$)       & 65.4$\pm$0.3 & 64.1$\pm$0.3           \\
     Age (Myr)                                  & \multicolumn{2}{c}{120$\pm$35} \\
     \hline
     \end{tabular}\\
      \footnotesize $^{a}$:SIMBAD Database, $^{b}$: Sung et al. (2013)
     \end{center}
    \end{table}

In order to analyze the kinematical properties of \object{HD 350731}, we used the system's center of mass velocity, distance and proper motion values. The proper motion data ($\mu_{\alpha}$ cos$\delta$, $\mu_{\delta}$)=(1.0$\pm$0.7, -8.7$\pm$0.8) mas yr$^{-1}$ were taken from the Fourth US Naval Observatory CCD Astrograph Catalog (UCAC4; Zacharias et al. 2013), whereas the center of mass velocity $V_\gamma=-10.1\pm0.4$ km s$^{-1}$ and distance d=703$\pm$34 pc were obtained in this study. The system's space velocity was calculated using Johnson \& Soderblom's (1987) algorithm. To obtain the space velocity precisely, first-order galactic differential rotation correction was taken into account (Mihalas \& Binney 1981). The differential rotation corrections were calculated as 16.23 and 0.89 km s$^{-1}$ and applied to $U$ and $V$ space velocity components, respectively. The $W$ velocity is not affected in this first-order approximation. As for the local standard of rest correction (LSR), Co\c{s}kuno\~{g}lu et al.'s (2011) values ($U$, $V$, $W$)=(8.50, 13.38, 6.49) km s$^{-1}$ were used and the total space velocity of \object{HD 350731} was obtained as S$_{tot}=15.30\pm$3.84 km s$^{-1}$. The corrected space velocity components are ($U$, $V$, $W$)=(6.10$\pm$2.39, -9.26$\pm$1.56, -10.54$\pm$2.57) km s$^{-1}$. The total space velocity and space velocity component values are in agreement with young-disc stars (Leggett, 1992).

To determine the population type of \object{HD 350731}, we used Dinescu, Girardi \& van Altena's (1999) N-body code and obtained the galactic orbit of the system. In this code, the timescale in generating the orbits was assumed to be 3 Gyr, and the calculation steps were 2 Myr. The 3 Gyr timescale was assumed so that precise orbits would be created, even though this is longer than the nuclear time scale of early-type stars. The orbits of \object{HD 350731} on the $X-Y$ and $X-Z$ planes around the galactic center are shown in Fig. 7. The system's apogalactic ($R_{max}$) and perigalactic ($R_{min}$) distances obtained were 7.72 and 7.09 kpc, respectively. According to N-body code, the maximum vertical separation from the galactic plane of the system is $\mid z_{max} \mid$=100 pc. The following formulae were used to derive the planar ($e_{p}$) and vertical ($e_{v}$) eccentricities:

\begin{equation}
e_p=\frac{R_{max}-R_{min}}{R_{max}+R_{min}},
\end{equation}

\begin{equation}
e_v=\frac{(|z_{max}|+|z_{min}|)}{R_m},
\end{equation}
where $R_{m}$ is the mean of $R_{min}$ and $R_{max}$. The planar and vertical eccentricities were calculated as $e_{p}=0.04$ and $e_{v}=0.01$, respectively. These eccentricities show that \object{HD 350731} is in a circular orbit around the mass center of the Galaxy and that it belongs to the young thin-disc population.

\begin{figure}
\begin{center}
\includegraphics[trim=0.5cm 0.5cm 0.5cm 0.5cm, clip=true, scale=0.50]{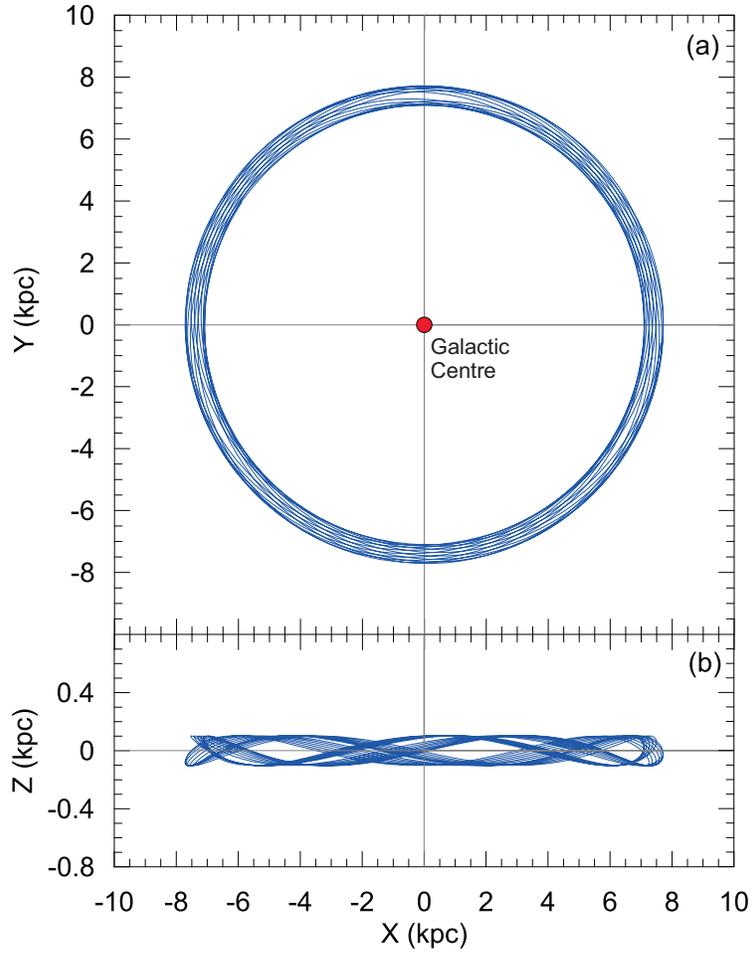}
\caption{Orbital motion of HD 350731 on projections of $X-Y$ (a)
and $X-Z$ (b) planes around galactic center for 3 Gyrs.}
\end{center}
\end{figure}

\section{Discussion and Conclusion}

Double-line detached eclipsing binaries (DBs) are valuable sources for determining the precise fundamental properties (mainly masses and radii) of stars. A recent catalogue of this type of binaries was published by Eker et al. (2014). It consists of 257 DBs; their 388 component stars have better than 5\% accuracy in their masses and radii. When one examines the basic properties of the DBs in the catalogue (especially mass ratio, mass, radius and temperatures of components), it can be seen that there is no detached eclipsing binary system which has absolute properties similar to \object{HD 350731}. The spectral type of the system was determined to be B8V+B8V, while the masses were derived to be $M_{1}=2.91\pm0.13$\,M$_{\odot}$ and $M_{2}=2.80\pm0.14$\,M$_{\odot}$ for the primary and secondary components, respectively. Therefore, DBs with mass ratio in the range of 0.95-1.0 and spectral type O and/or B need to be further studied in order to fill the gap in this parameter range. This is necessary to understand the evolution and structure of early-type stars.

The mass-luminosity relation (MLR) was updated recently by Eker et al. (2015) using DBs data. In this study, they identified four mass domains and derived MLRs for these mass ranges. For comparison, using the MLR given in the mass range of 2.4\,M$_{\odot}$ and 7\,M$_{\odot}$, we calculated luminosity values on a logarithmic scale to be 1.96\,L$_{\odot}$ and 1.89\,L$_{\odot}$, for the primary and secondary components, respectively. The predicted values from the updated classical MLR agree with the values derived from Stefan-Boltzmann law (see Table 8).

Spectral data enables us to find temperatures and projected rotational velocities of the components. For the model atmosphere application, the disentangled spectra of the components obtained from KOREL analysis were used. As a results the temperature of the primary and secondary components was found to be 12000 K and 11830 K, respectively. The surface gravity values ($\log g_{1,2}$) calculated from the masses and radii of the components were adopted during spectral analysis. The projected rotational velocities ($v\sin i$) of the components were measured to be 69.2 km\,s$^{-1}$ and 70.1 km\,s$^{-1}$ for the primary and secondary components, respectively. Within errors the measured $v\sin i$ values agree with the pseudo-synchronous velocities of the components in Table 8, which were calculated on the basis of the formulations by Hut (1981).

The orbit of the system is slightly eccentric determined from the orbital solution and also an analysis of the multi-color light curves and radial velocities of the components. The secondary minimum can be seen to have shifted from the orbital phase of 0.5. This is an indication of apsidal motion in the system. We collected 15 minima times together with newly-measured eclipse times to study the apsidal motion by means of $O-C$ analysis. The apsidal motion parameters could not be determined accurately since the observed eclipse times were not-well covered. The apsidal motion was found at a rate of $\dot{\omega}=0.0170\pm0.0082$ degree cycle$^{-1}$, which corresponds to an apsidal motion period of about 92 yr.

The locations of the components of \object{HD 350731} are seen in the Hertzsprung-Russell (HR) diagram and plane of $M-R$ in Fig. 8. Zero Age Main Sequence (ZAMS) and the evolutionary tracks for the exact masses of the components, and isochrones for solar chemical composition, are taken from the Yonsei--Yale (Y$^{2}$) series of Yi et al. (2001). As seen in Fig. 8, the components are a bit away from the ZAMS line and the age of the system was estimated to be 120$\pm$35 Myr from Y$^{2}$ isochrones. Kinematic analysis indicated that the system leaves the galactic plane just about 100 pc during its movement on galactic orbit. This is evidence for membership of the thin-disc population by \object{HD 350731}. The positions of the components in the HR diagrams are in agreement with the evolutionary tracks for the masses of 2.91\,M$_{\odot}$ and 2.80\,M$_{\odot}$.

As a general conclusion, we can mention that the photometric and spectroscopic analysis of \object{HD 350731} leads us to extend the database of DBs with intermediate mass components having similar properties. This is important since in the catalogue of DBs by Eker et al. (2014) there are only a few systems with a mass ratio close to $q=1.0$ and components having masses greater than 2.8\,M$_{\odot}$ (e.g. $\eta$ Mus, V799 Cas, V906 Sco). The unevolved binary system \object{HD 350731} belongs to the young thin-disc population of the Galaxy with an estimated age of 120$\pm$35 Myr. In order to make abundance analysis in detail, and also verify population type of the system, more high resolution spectroscopic data is needed. New eclipse times are also required to enlarge the time interval for better analysis and to confirm the apsidal motion parameters.

\begin{figure}
\begin{center}
\includegraphics[trim=0.5cm 0.5cm 0.5cm 0.5cm, clip=true, scale=0.70]{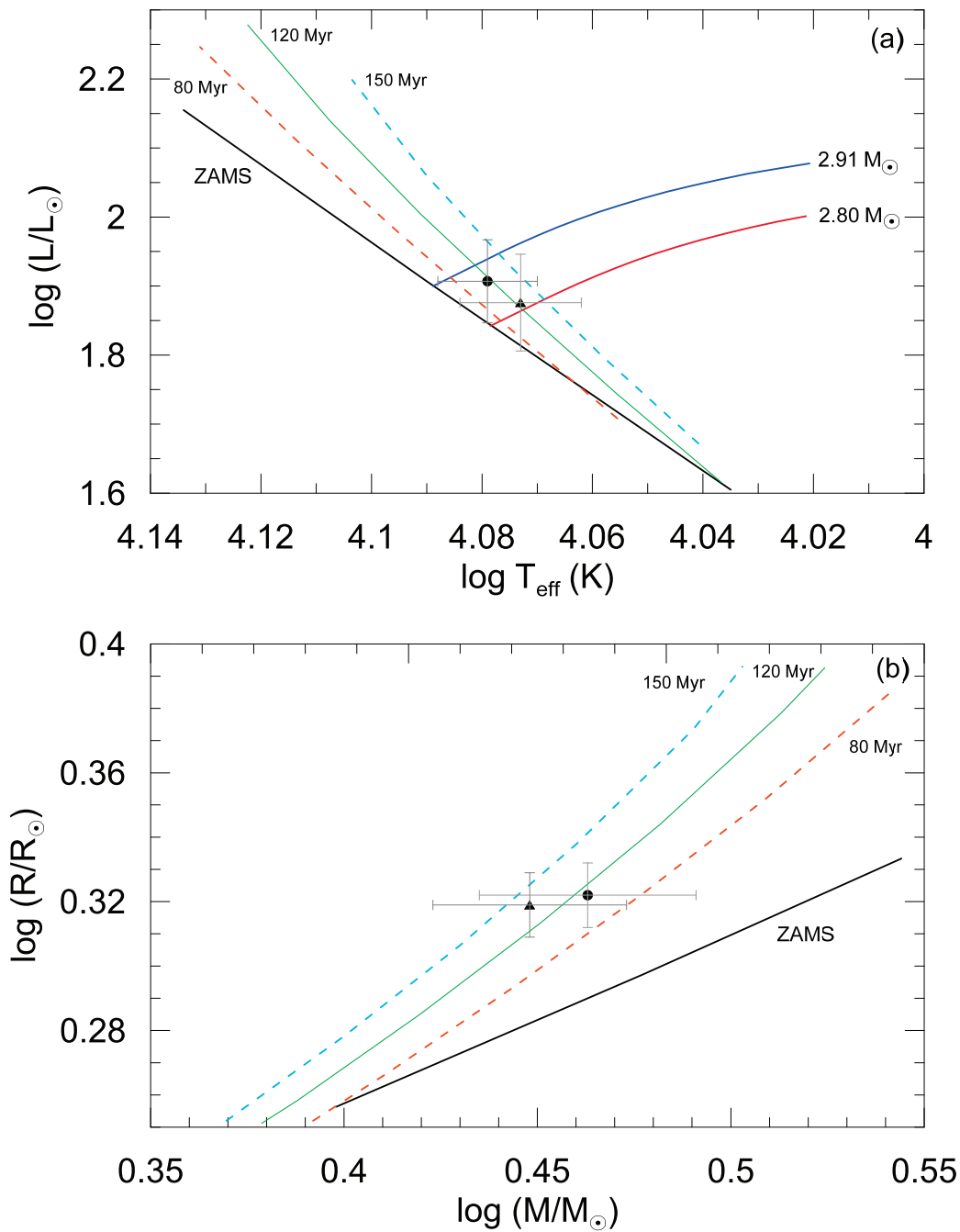}
\caption{Locations of primary and secondary components
of HD 350731 in log \emph{L} - log \emph{T$_{eff}$} (a) and $\log M$-$\log R$ planes (b). Evolutionary tracks for masses of 2.91\,$M_{\odot}$ and 2.80\,M$_{\odot}$ (solid lines), isochrones for ages of 80 Myr, 120 Myr and 150 Myr (dashed lines), and ZAMS for solar chemical composition adopted from Yi et al. (2001).}
\end{center}
\end{figure}




\acknowledgments

This research was supported by the Scientific and Technological Research Council of Turkey (T\"UB\.ITAK, Grant no. 111T224). The authors would like to thank the anonymous referee for valuable suggestions and comments which helped us to improve the study.  We thank \c{C}anakkale Onsekiz Mart University Astrophysics Research Center and Ulup{\i}nar Observatory together with \.{I}stanbul University Observatory Research and Application Center for their support and allowing use of T122 and IST60 telescopes. The project was supported partly by National Planning Agency (DPT) of Turkey (project DPT-2007K120660 carried out at \c{C}anakkale Onsekiz Mart University) and the Scientific Research Projects Coordination Unit of Istanbul University (project no. 3685). We gratefully acknowledge the support of the NRC (Canada) Herzberg Institute of Astrophysics. The authors would especially like to thank Dr. D. Bohlender and Dr. D. Monin for their hospitality and allowing telescope time for observations at the Dominion Astrophysical Observatory, Canada. This research has made use of the SIMBAD and NASA Astrophysics Data System Bibliographic Services.

\end{document}